\documentclass[12pt,preprint]{aastex}

\usepackage{amsmath}
\usepackage{graphicx}
\usepackage{comment}
\usepackage{color} 
\newcommand{\fig}[1]{Fig.~\ref{#1}}

\newcommand{\sect}[1]{Section~\ref{#1}}

\begin{document}
\title{Eruption of Solar Magnetic Flux Ropes Caused by Flux Feeding}
\author{Quanhao Zhang\altaffilmark{1,2,3,4}, Yuming Wang\altaffilmark{1,3,4}, Rui Liu\altaffilmark{1,3,5}, Jie Zhang\altaffilmark{6}, Youqiu Hu\altaffilmark{1}, Wensi Wang\altaffilmark{1,3,4}, Bin Zhuang\altaffilmark{1,4}, Xiaolei Li\altaffilmark{1,4}}
\altaffiltext{1}{CAS Key Laboratory of Geospace Environment, Department of Geophysics and Planetary Sciences, University of Science and Technology of China, Hefei 230026, China}
\altaffiltext{2}{State Key Laboratory of Space Weather, Chinese Academy of Sciences, Beijing 100190}
\altaffiltext{3}{CAS Center for Excellence in Comparative Planetology, Hefei 230026, China}
\altaffiltext{4}{Mengcheng National Geophysical Observatory, School of Earth and Space Sciences, University of Science and Technology of China, Hefei 230026, China}
\altaffiltext{5}{Collaborative Innovation Center of Astronautical Science and Technology, Hefei, Anhui 230026, China}
\altaffiltext{6}{Department of Physics and Astronomy, George Mason University, 4400 University Drive, MSN 3F3, Fairfax, VA 22030, USA}
\email{zhangqh@ustc.edu.cn}

\begin{abstract}
Large-scale solar eruptions are believed to have a magnetic flux rope as the core structure. However, it remains elusive as to how the flux rope builds up and what triggers its eruption. Recent observations found that a prominence erupted following multiple episodes of ``flux feeding". During each episode, a chromospheric fibril rose and merged with the prominence lying above. In this letter, we carried out 2.5-dimensional magnetohydrodynamic (MHD) numerical simulations to investigate whether the flux-feeding mechanism can explain such an eruption. The simulations demonstrate that the discrete emergence of small flux ropes can initiate eruptions by feeding axial flux into the preexistent flux rope until its total axial flux reaches a critical value. The onset of the eruption is dominated by an ideal MHD process. Our simulation results corroborate that the flux feeding is a viable mechanism to cause the eruption of solar magnetic flux ropes.
\end{abstract}


\section{Introduction}
\label{sec:introduction}
Large-scale solar eruptions are manifested as the observed phenomena of flares, prominence/filament eruptions, and coronal mass ejections (CMEs). It is widely accepted that these kinds of events are intimately associated with a coronal magnetic flux rope system and are essentially different manifestations of the same physical process, i.e., the eruption of the rope system \citep{Zhang2001,vanDriel2015a,Green2018,Jiang2018,Yan2018}. A typical scenario is that, during a flux rope eruption, the prominence/filament contained in the rope also erupts with the rope, and the magnetic reconnection in the current sheet formed beneath the rope dramatically converts free magnetic energy in the coronal magnetic system into thermal energy and non-thermal particle acceleration, so that a flare occurs; this flux rope further propagates outwards and expands, so as to be observed as a CME in the corona and the interplanetary space \citep[e.g.,][]{Lin2000a}. These large-scale eruptive activities are generally considered to be the major disturbance affecting the solar-terrestrial system \citep[e.g.,][]{Shen2014}. Therefore, it has great significance to study the formation process of an erupting magnetic flux rope and its trigger mechanism.
\par
Various theoretical models have been proposed to investigate the eruptive mechanism of flux ropes, either based on magnetic reconnection \citep{Antiochos1999a,Chen2000a,Moore2001a,Sterling2004,Archontis2008b,Inoue2015} or ideal MHD instabilities \citep{Romano2003,Torok2003a,Kliem2006a,Fan2007a,Aulanier2010a,Guo2010,Savcheva2012b}.
It was also suggested by many authors that catastrophes could be responsible for solar eruptions: the onset of the eruption corresponds to a catastrophic loss of equilibrium \citep[][]{Forbes1991a,Isenberg1993a,Lin2001a,Chen2007a,Demoulin2010a,Longcope2014a,Kliem2014}. Flux rope catastrophes could be triggered by various physical processes. For example, it was found that there exists a critical value of the total axial (also called toroidal in a Tokamak configuration) magnetic flux of a flux rope \citep[e.g.][]{Bobra2008,Su2009,Su2011a,Zhang2016a,Zhang2017a,Zhang2017,Zhuang2018}. If the axial flux of the rope is smaller than this critical value, the rope system stays in equilibrium states; when this critical value is exceeded, loss of equilibrium occurs in the rope system: the flux rope jumps upward, with magnetic reconnection occurring below it, so that the rope erupts outward. This critical axial flux is of the order $10^{19}\sim10^{20}$ Mx, and is influenced by various conditions, such as photospheric magnetic flux distributions \citep[e.g.,][]{Zhang2017a}. 
\par
\begin{figure*}
\includegraphics[width=\hsize]{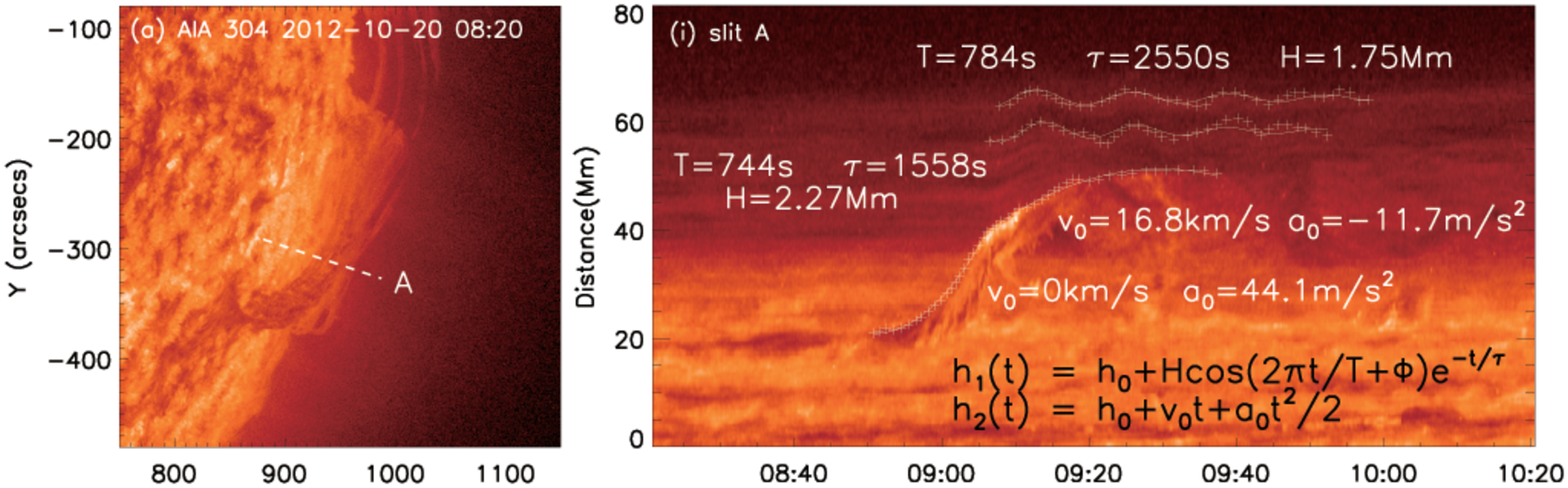}
\caption{The observations of a typical flux feeding process. The left panel is the Atmospheric Imaging Assembly (AIA) 304 $\mathrm{\AA}$ observation of the prominence; the right panel is the slice-time plot along the slit marked as ``A" in the left panel. This figure is adapted from \cite{Zhang2014}.}\label{fig:obs}
\end{figure*}

Recently, it was observed by \cite{Zhang2014} (hereafter Paper I) that a sequence of flux feeding episodes occurred within the two-day period prior to the eruption of a prominence. As shown in the right panel in \fig{fig:obs}, a chromospheric fibril appeared as a dark structure at about 20 Mm along the slit before about 08:50 UT, after which it rose and merged with the prominence within about 40 $\sim$ 60 Mm along the slit. During a flux feeding process, magnetic flux and mass are injected into the target prominence from the chromosphere underneath. This is reminiscent of bubbles rising and expanding into quiescent prominences \citep[e.g.,][]{Berger2010}, as well as the transfer of magnetic flux and current between the different branches in a double-decker configuration \cite[e.g.,][]{Liu2012b,Kliem2014a,Cheng2014}. As observed in Paper I, flux feeding events successively occurred 3 times, which increased the slow-rising velocity of the prominence. Eventually, the prominence erupted. Therefore it was suggested that the eruption could be initiated by flux feeding processes. This also indicates that flux feeding could be regarded as one of the precursors of solar eruptions \cite[e.g.][]{Wang2017a}. The physical nature of the flux rope eruptions initiated by flux feeding, however, remains unclear; there are still many issues about this scenario. The most prominent one is, why could flux feeding cause the prominence to erupt? Is this merely a coincidence or actually an indication of some physical mechanism? Moreover, there were 3 flux feeding episodes in the pre-eruptive phase of the prominence. Why did the prominence not erupt after the 1$^{st}$ and the 2$^{nd}$ flux feeding episodes, but only erupt after the 3$^{rd}$ one? Whether it was the three flux feeding episodes as a whole or only the 3$^{rd}$ one that is responsible for the onset of the eruption? These questions could hardly be resolved based on observational results alone. Theoretical investigations are needed to shed light on the physical nature of the flux rope eruptions caused by flux feeding processes.
\par
In this letter, we carry out numerical simulations to investigate the physical nature of the flux rope eruption initiated by flux feeding. The major science question is about the influence of flux feeding processes on coronal flux rope systems, especially the role that flux feeding plays in the onset of the eruptions.  The rest of the paper is arranged as follows: the simulating procedures are introduced in \sect{sec:method}; simulation results of a typical flux feeding event are presented in \sect{sec:result}; the physical nature of the onset of the eruptions is investigated in \sect{sec:analysis}. Finally, discussion and conclusion are given in \sect{sec:dc}.

\section{Simulating procedures}
\label{sec:method}
For 2.5-dimensional cases (with $\partial/\partial z=0$) in Cartesian coordinates, the magnetic field can be denoted as
\begin{align}
\textbf{B}=\triangledown\psi\times\hat{\textbf{\emph{z}}}+B_z\hat{\textbf{\emph{z}}},\label{equ:mf}
\end{align}
where $\psi$ is the magnetic flux function, $B_z$ is the component of $\textbf{B}$ in $z-$direction. Basic equations and procedures to obtain the initial state are introduced in Appendix \ref{ape:equations}. The background field is a partially open bipolar field (\fig{fig:feeding}(a)). Anomalous resistivity is used here so that magnetic reconnection is restricted within the region of current sheets:
\begin{align}
\eta=
\begin{cases}
0,& ~j\leq j_c\\
\eta_m\mu_0v_0L_0(\frac{j}{j_c}-1)^2.& ~j> j_c \label{equ:res}
\end{cases}
\end{align}
Here $\eta_m=10^{-4}$ and $L_0=10^7$ m, and $v_0=\sqrt{RT_0}=128.57$ km s$^{-1}$, where $T_0=10^6\mathrm{~K}$; $R=1.65\times10^4$ J kg$^{-1}$ K$^{-1}$ is the specific gas constant and $\mu_0$ is the vacuum magnetic permeability; the critical current density is $j_c=2.37\times10^{-4}$ A m$^{-2}$. 
\par
The initial state is a stable equilibrium state (\fig{fig:feeding}(b)): 
\begin{figure*}
\includegraphics[width=\hsize]{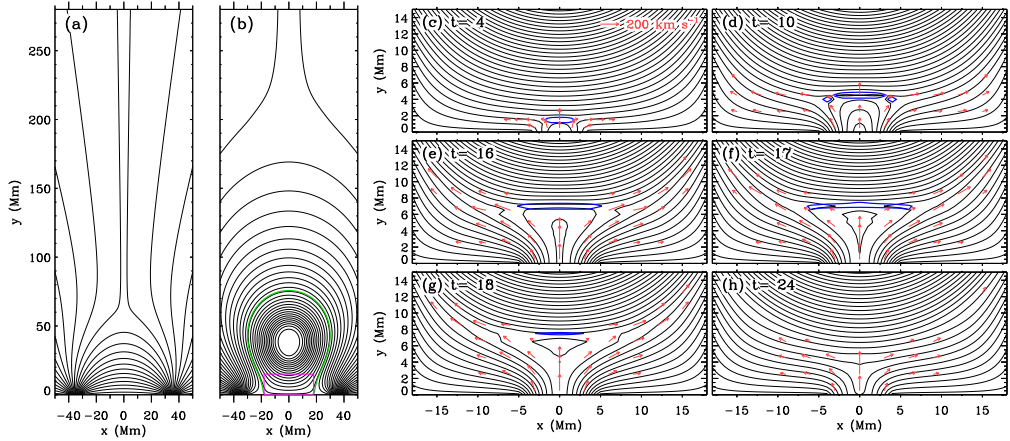}
\caption{Simulation results of a flux feeding process with $C_E=$1.90. Panel (a) and (b) show the magnetic configurations of the background field and the initial state, respectively; the green curve marks the boundary of the rope; the pink box represents the region illustrated in Panels (c)-(h). The black curves in panels (c)-(h) are the magnetic field lines; the blue curves are the contours of the current density $j=5.63\times10^{-4}$ A m$^{-2}$. The red arrows illustrate the distribution of the velocity in x-y plane; the length of arrows are proportional to the velocities; an example of 200 km s$^{-1}$ is plotted in panel (c). The time is in the unit of $\tau_A$.}\label{fig:feeding}
\end{figure*}
a flux rope is embedded in the bipolar background field. In our simulation, the rising fibril in the scenario of flux feeding is represented by a small flux rope, which emerges from below the pre-existing flux rope, rises and interacts with the pre-existing rope. For simplicity, the pre-existing large flux rope is called major flux rope hereafter. Assume that the small rope, whose radius is $a=5$ Mm, begins to emerge at $t=0$ in the central region of the base right below the major rope, and the emergence ends at $t=\tau_E=30\tau_A$ s; $\tau_A=L_0^2\sqrt{\mu_0\rho_0}/\psi_0 =17.4$ s is the characteristic Alfv\'{e}n transit time, where $\rho_0=3.34\times10^{-13}\mathrm{~kg~m^{-3}}$ and $\psi_0=3.73\times10^3\mathrm{~Wb~m^{-1}}$. With a constant emerging speed, the emerged part of the small rope at time $t$ is located within $-x_E\leqslant x\leqslant x_E$, where
\begin{align}
x_E=(a^2-h_E^2)^{1/2}, h_E=a(2t/\tau_E-1).
\end{align}
Based on this, the emergence of the small flux rope is achieved by adjusting $\psi$, $B_z$, the velocities $v_{x,y,z}$, the temperature $T$, and the density $\rho$ at the base of the emerged part of the small rope ($y=0, -x_E\leqslant x\leqslant x_E$):
\begin{align}
&\psi(t,x,y=0)=\psi_i(x,y=0)+\psi_E(t,x),\\
&\psi_E(t,x)=\frac{C_E}{2}\mathrm{ln}\left(\frac{2a^2}{a^2+x^2+h_E^2}\right)\label{equ:psi},\\ 
&B_z(t,x,y=0)=C_Ea(a^2+x^2+h_E^2)^{-1}\label{equ:bz},\\
&v_y(t,x,y=0)=v_E=2a/\tau_E,~v_x(t,x,y=0)=v_z(t,x,y=0)=0,\\
&T(t,x,y=0)=2\times10^5\mathrm{~K},~\rho(t,x,y=0)=1.67\times10^{-12}\mathrm{~kg~m^{-3}}.
\end{align}
Here $\psi_i$ is the magnetic flux function of the initial state. Apart from during the emergence of the small rope, $\psi$ at the base is fixed at $\psi_i$, so that it corresponds to the photosphere. $B_z$ is positive and $\mathrm{\textbf{B}}_{xy}$ (the component of \textbf{B} in $x$-$y$ plane) is counterclockwise in both the small and the major ropes. It is widely accepted that the distribution of coronal magnetic field plays a dominant role in how the eruption of a flux rope is triggered \citep[e.g.,][]{Sun2012a}. Thus the influence of flux feeding on the major rope should be sensitive to the scale of the strength of magnetic field in the emerging small rope. As shown in Equation (\ref{equ:psi}) and Equation (\ref{equ:bz}), the parameter $C_E$ determines the magnetic field strength of the small rope; its dimensionless values quoted in the rest of the paper are given in the unit of $\psi_0=3.73\times10^3\mathrm{~Wb~m^{-1}}$. In our simulations, we change $C_E$ to investigate the influences of different flux feeding processes on the major flux rope system. It should be noted that the science focus and simulating procedures in this work are quite different from those of \cite{Zhang2017a}, in which the catastrophic behaviors of a single flux rope was investigated.
\par

\section{Simulation results}
\label{sec:result}
\begin{figure*}
\includegraphics[width=\hsize]{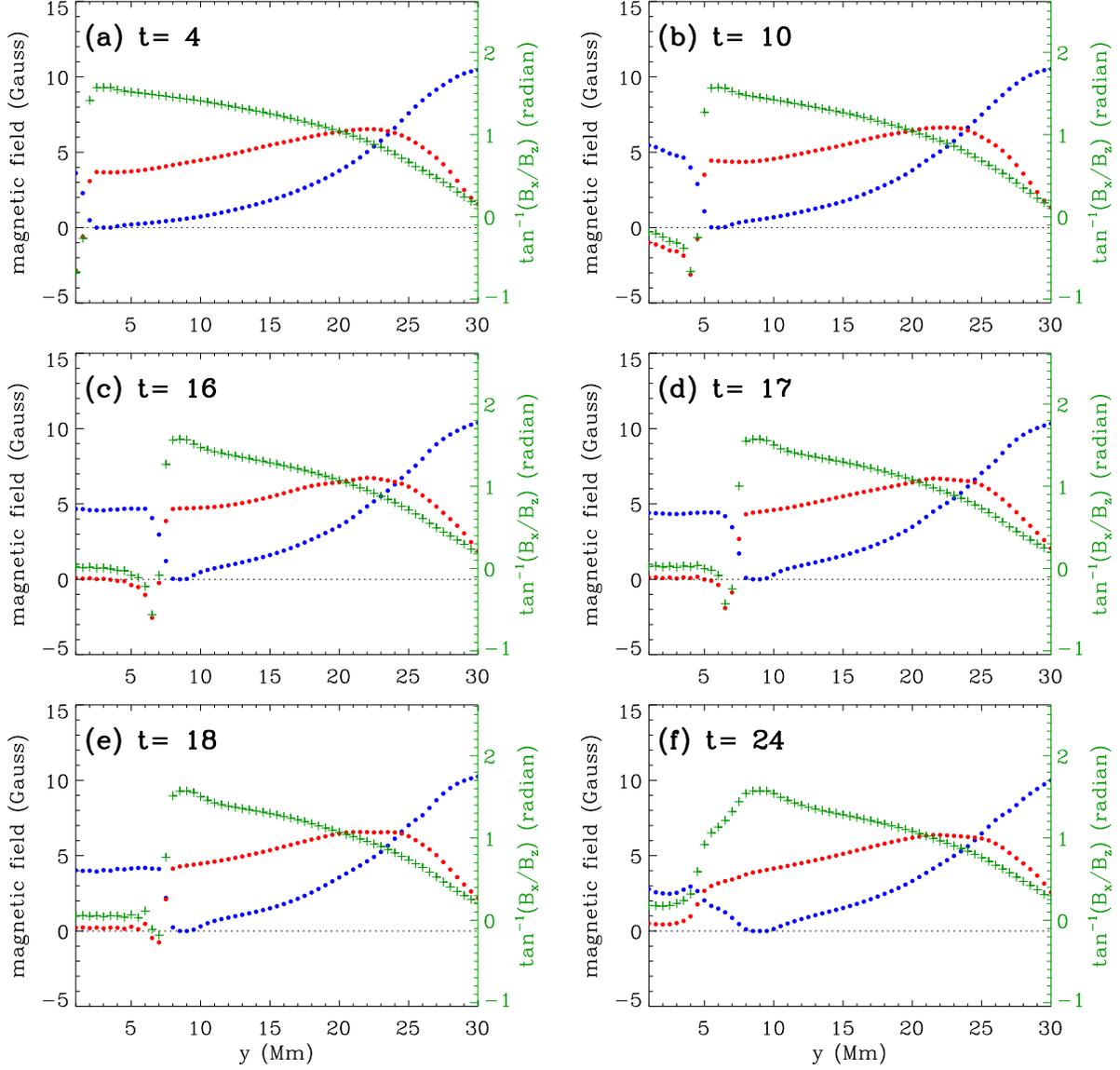}
\caption{The distributions of $B_x$ (red dots), $B_z$ (blue dots), and $\tan^{-1}(B_x/B_z)$ (green plus signs) along $x=0$ in the states shown in \fig{fig:feeding}(c) to \ref{fig:feeding}(h). The black horizontal dotted lines represents $B=0$. The time is in the unit of $\tau_A$.}\label{fig:magdis}
\end{figure*}
\begin{figure*}
\includegraphics[width=\hsize]{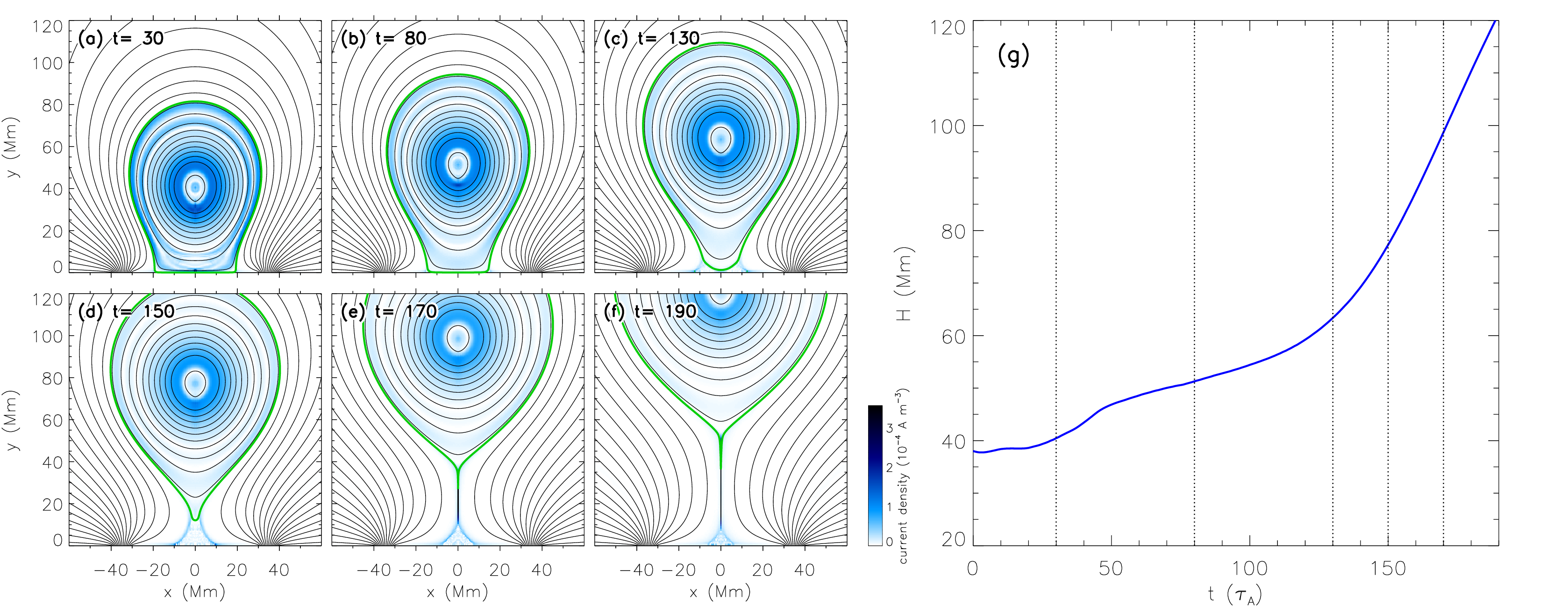}
\caption{The eruptive process of the case with $C_E=$1.90. The blue curve in panel (g) is the evolutionary profile of the height of the rope axis, $H$. Panels (a)-(f) show central sections of the domain during the evolution, in which the blue color depicts the distribution of the current density, and the green curves mark the boundary of the rope. The corresponding times of the states shown in panels (a)-(h) are marked by the vertical dotted lines in panel (g).}\label{fig:erwr}
\end{figure*}

The simulation result of a typical flux feeding process with $C_E=$1.90 is shown in \fig{fig:feeding}. At the early stage of the flux feeding process, the emerging small flux rope appears below the major rope, as shown in \fig{fig:feeding}(c). A horizontal current sheet forms at the interface between the small and major ropes, as marked by the blue curves in \fig{fig:feeding}, which are the contours of the current density $j=5.63\times10^{-4}$ A m$^{-2}$. The emerged small rope could be clearly recognized in \fig{fig:feeding}(g) and the corresponding distribution of $B_x$ in \fig{fig:magdis}(e). As a result of the magnetic reconnection within the current sheet, the magnetic field lines of the small rope gradually reconnect with those of the major rope (see \fig{fig:feeding}(e)-\ref{fig:feeding}(h)). The height of the current sheet gradually increases with time, triggering flows within the major rope, as illustrated by the red arrows in \fig{fig:feeding}(e)-\ref{fig:feeding}(h). Eventually, the two flux ropes merge together. Note that since the major rope sticks to the photosphere, the reconnection occurs immediately after the small rope begins to emerge. The topology of the resultant flux rope system after flux-feeding is shown in \fig{fig:erwr}(a). 
\par
Further evolution of the resultant flux rope system indicates that this flux feeding process with $C_E=$1.90 eventually triggers the major rope to erupt. As shown in \fig{fig:erwr}(g), the eruption of the rope occurs after the flux feeding process. After the onset of the eruption, the lower boundary of the rope is not detached from the photosphere instantly, but keeps sticking to the photosphere for a certain period (\fig{fig:erwr}(b)-\ref{fig:erwr}(c)). As the height of the rope increases, the lower part of the rope, along with the adjacent background field lines, are stretched, during which the flux rope is gradually accelerated. Eventually, a vertical current sheet forms beneath the flux rope, as shown in \fig{fig:erwr}(d)-\ref{fig:erwr}(f). The magnetic reconnection that occurs in this current sheet should drive the further acceleration of the flux rope. The obvious delay of the appearance of this current sheet relative to the onset of the eruption indicates that the eruption should be triggered by an ideal process. This is consistent with the observations in Paper I, in which there was no intense heating around the source region of the prominence during the early period of its eruption, indicating that fast magnetic reconnection plays no crucial role in triggering the eruption. It is noteworthy that the initial state is a stable equilibrium: if there is no flux feeding process, the major rope will keep sticking to the photosphere forever.

\section{Analysis}
\label{sec:analysis}
As shown in the simulations demonstrated in \sect{sec:result}, flux feeding is able to eventually cause a flux rope system to erupt, consistent with the conclusion in Paper I. To further understand the physical nature of this scenario, detailed investigation about the influence of different settings of the flux feeding is needed. It has already been mentioned in \sect{sec:method} that $C_E$ determines the magnetic field strength in the emerging small flux rope, so that cases with different $C_E$ correspond to different intensities of flux feeding.
\par
The properties of the major rope is characterized by the axial magnetic flux, $\Phi_z$, and the poloidal magnetic flux per unit length along the $z-$direction, $\Phi_p$. In the initial state shown in \fig{fig:feeding}(b), the axial flux $\Phi_{z0}=9.31\times10^{19}~\mathrm{Mx}$, the poloidal flux $\Phi_{p0}=1.49\times10^{10}~\mathrm{Mx}~\mathrm{cm}^{-1}$. Assuming the length of the rope is 100 Mm, the total poloidal flux of the flux rope is of the order $1.5\times10^{20}$ Mx. For the case with $C_E=1.90$ shown in \sect{sec:result}, the axial flux $\Phi_z$ of the resultant rope at $t=30\tau_A$ increases to $11.73\times10^{19}~\mathrm{Mx}$, whereas the poloidal flux $\Phi_p$ is still $1.49\times10^{10}~\mathrm{Mx}~\mathrm{cm}^{-1}$, almost the same as the initial state. This indicates that the twist angle within the rope should decrease after flux feeding. Simulation results with other different $C_E$ also comes to the similar conclusion, indicating that flux feeding processes only inject axial flux into the major rope. This is because the poloidal flux of the small rope is entirely cancelled out by the magnetic reconnection during its merging process with the major rope. The injected axial flux is mainly distributed near the boundary of the flux rope, which results in the current in this region after flux feeding (see \fig{fig:erwr}(a)).
\par
\begin{figure*}
\centering
\includegraphics[width=0.8\hsize]{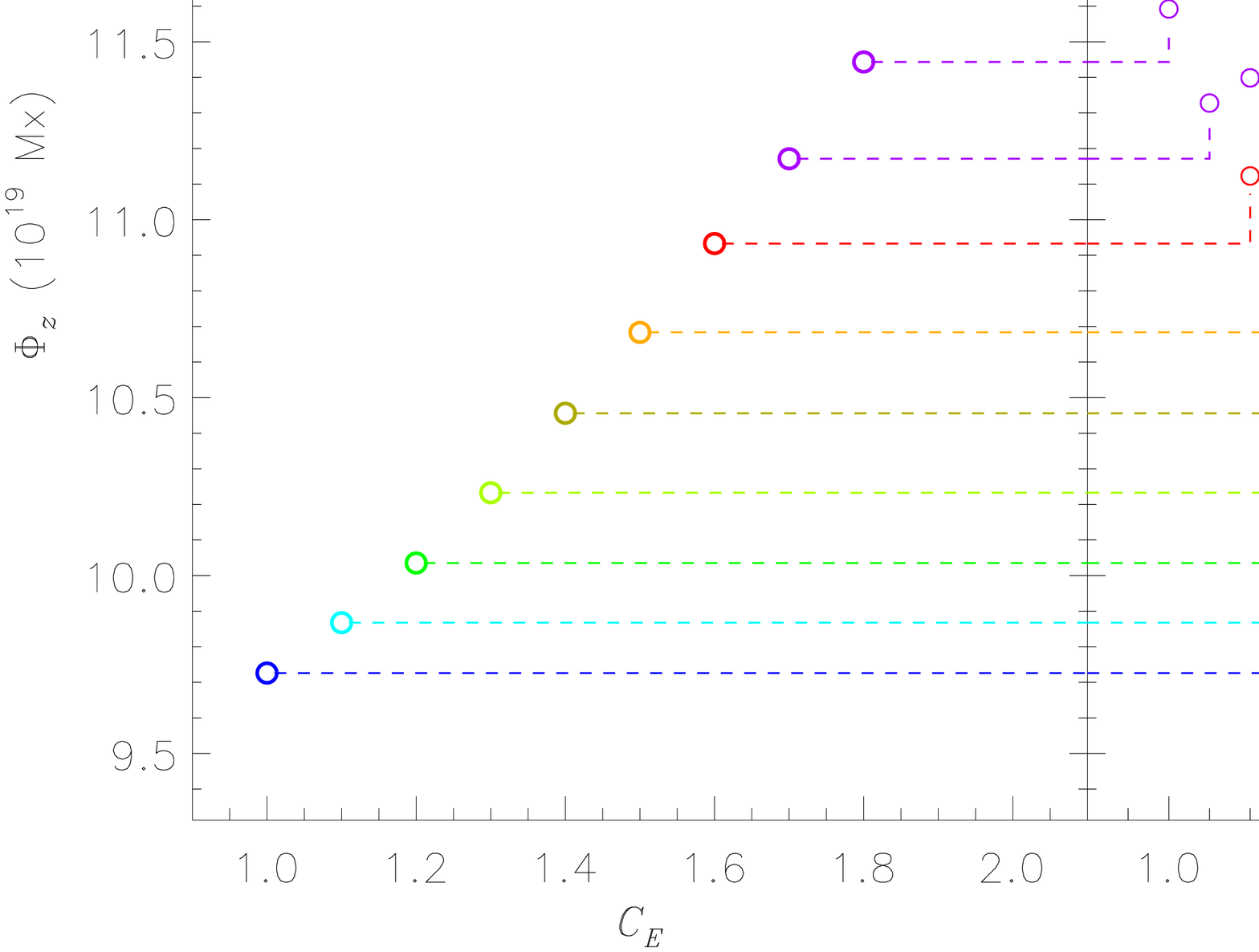}
\caption{Total axial flux $\Phi_z$ of the resultant rope at $t=30\tau_A$ for different $C_E$. The eruptive cases are plotted in solid dots, and non-eruptive cases in circles. Panel (a) shows $\Phi_z$ of the resultant rope after the 1$^{st}$ round of flux feeding, and non-eruptive cases with different $C_E$ are in different colors. Panel (b) shows $\Phi_z$ after the 2$^{nd}$ round, and their colors are the same as their corresponding pre-feeding states. The dashed lines do not have many physical implications, but mark the correspondence between the pre- and post-feeding states.}\label{fig:crit}
\end{figure*}

The flux feeding process, however, is not always able to trigger the major flux rope to erupt; it requires certain threshold. For the cases with different $C_E$, $\Phi_z$ of the resultant rope at $t=30\tau_A$ is plotted in \fig{fig:crit}(a); the non-eruptive cases (i.e. the major rope keeps sticking to the photosphere after  flux feeding) with different $C_E$ are plotted in circles with different colors, while the eruptive ones in black solid dots. For the case with larger $C_E$, the magnetic field in the small emerging rope is stronger, so that more axial flux is injected. It is obvious in \fig{fig:crit}(a) that $\Phi_z$ of the resultant rope in the eruptive cases is larger than that in the non-eruptive ones. For each non-eruptive case in \fig{fig:crit}(a), through using the non-eruptive state as the new pre-feeding state, we let a new small rope emerge from below the major rope, and these cases are called the $2^{nd}$ round of flux feeding; the corresponding $\Phi_z$ at $t=30\tau_A$ is plotted in \fig{fig:crit}(b), and their colors are the same as their corresponding pre-feeding states. Similarly, the non-eruptive cases are plotted in circles, and eruptive cases in dots. For clarification, the cases starting from the initial state in \fig{fig:crit}(a) are called the $1^{st}$ round. It is demonstrated in \fig{fig:crit} that the eruptive and non-eruptive cases are separated. There should exist a critical value $\Phi_{zc}$ of the order $1.2\times10^{20}~\mathrm{Mx}$. If $\Phi_z<\Phi_{zc}$ (circles), the eventual height of the major rope is finite, whereas if $\Phi_z\geq\Phi_{zc}$ (dots), the eventual height should be infinite.
\par

\section{Discussion and conclusion}
\label{sec:dc}
In this letter, we have carried out MHD numerical simulations to investigate the effect of flux feeding on coronal flux rope systems. In our simulations, it is found that the flux feeding processes only inject axial magnetic flux into the major rope, whereas the poloidal magnetic flux of the rope remains almost unchanged. The physical scenario of the eruption caused by flux feeding is: by injecting axial flux into a flux rope in an incremental and intermittent fashion, flux feeding effectively drives the rope to evolve toward a critical condition and eventually can trigger its eruption. Therefore, our simulation results corroborate that flux feeding is a viable mechanism to cause the eruption of solar magnetic flux ropes. For the major flux rope, there exists a threshold $\Phi_{zc}$: if the major rope's axial flux $\Phi_z$ is below $\Phi_{zc}$, it will keep sticking to the photosphere, no matter how many flux feeding episodes have occurred; on the other hand, if $\Phi_z$ exceeds $\Phi_{zc}$, the rope system will erupt. 
\par
The existence of the threshold $\Phi_{zc}$ indicates that the number of flux-feeding episodes is not important; only when the amount of its axial flux exceeds the critical value will the major rope erupt. Based on this result, the evolution of the observational event analyzed in Paper I can be interpreted as follows. The injected axial magnetic flux via the $1^{st}$ and $2^{nd}$ observed flux feeding episodes might not be sufficient for the flux rope embedding the prominence to reach its critical state, thus the prominence remained in a quasi-equilibrium (slow rising) state. But the 3rd episode of flux feeding became the ``last straw",  so that the flux rope erupted. The early flux feeding processes might not trigger the eruption, but with each episode of flux feeding the rope system was one step closer to the eruption. 
\par
As introduced in \sect{sec:introduction}, previous studies suggested the presence of a critical axial flux for flux ropes; a catastrophe occurs if this critical value is reached. Our simulation results also support this theoretical conclusion. In the eruptions caused by flux feeding, the flux feeding processes continually inject axial flux, acting as a build-up towards the onset of the eruption. When the critical axial flux is reached, an upward catastrophe is triggered, and the further evolution of the upward catastrophe, along with the magnetic reconnection within the current sheet below the rope, drives the eruption of the flux rope  (see \cite{Green2018} for the classification of ``trigger" and ``driver" for solar eruptions). The critical value in our simulation is of the order $1.2\times10^{20}$ Mx, which is comparable with both the derived values in theoretical analyses \citep[e.g.][]{Su2011a,Zhang2017a} and the observed magnetic fluxes of CMEs \citep[e.g.,][]{Qiu2007,Wang2015a,Hu2015,Wang2017,Gopalswamy2017}. The increase of $\Phi_z$ in the theoretical studies mentioned above is artificial, i.e. the discovered upward catastrophe is only a phenomenon in the parameter space of $\Phi_{z}$, not reflecting the dynamic evolution of the system. The flux rope eruption in the corona, however, is a dynamic phenomenon in the physical space. Our simulations demonstrate flux feeding as a viable mechanism to prepare a flux rope for the upward catastrophe in the physical space, not just in the parameter space. It is noteworthy that, because of the different initial states and simulating procedures, the critical axial fluxes in, e.g., \cite{Zhang2017a}, are not exactly the same as ours, and the rope systems eventually reach equilibria in that study.
\par
This research is supported by the National Natural Science Foundation of China (NSFC 41804161, 41774178, 41761134088, 41774150, 41842037 and 41574165), the Strategic Priority Program of CAS (XDA15017300 and XDB41000000), and the fundamental research funds for the central universities. This project is also supported by the Specialized Research Fund for State Key Laboratories. We acknowledge for the data resources from National Space Science Data Center, National Science \verb"&" Technology Infrastructure of China (www.nssdc.ac.cn). The authors thank prof. Jun Lin for his valuable comment on the simulating procedures. The authors also thank the anonymous referee for his/her comments on the analysis of the data.

\appendix

\section{Basic equations and initial preparations}
\label{ape:equations}
Through using Equation \ref{equ:mf}, the 2.5-Dimensional MHD equations can be rewritten in dimensionless form as follows:
\begin{align}
&\frac{\partial\rho}{\partial t}+\triangledown\cdot(\rho\textbf{\emph{v}})=0,\label{equ:cal-st}\\
&\frac{\partial\textbf{\emph{v}}}{\partial t}+\textbf{\emph{v}}\cdot\triangledown\textbf{\emph{v}}+\triangledown T +\frac{T}{\rho}\triangledown\rho+\frac{2}{\rho\beta_0}(\vartriangle\psi\triangledown\psi+B_z\triangledown B_z+\triangledown\psi\times\triangledown B_z)+g\hat{\textbf{\emph{y}}}=0,\\
&\frac{\partial\psi}{\partial t}+\textbf{\emph{v}}\cdot\triangledown\psi-\frac{2\eta}{\beta_0}\vartriangle\psi=0,\\
&\frac{\partial B_z}{\partial t}+\triangledown\cdot(B_z\textbf{\emph{v}})+(\triangledown\psi\times\triangledown v_z)\cdot\hat{\textbf{\emph{z}}}-\frac{2\eta}{\beta_0}\vartriangle B_z=0,\\
&\frac{\partial T}{\partial t}+\textbf{\emph{v}}\cdot\triangledown T +(\gamma-1)T\triangledown\cdot\textbf{\emph{v}}-\frac{4\eta(\gamma-1)}{\rho R\beta_0^2}\left[(\vartriangle\psi)^2+|\triangledown\times(B_z\hat{\textbf{\emph{z}}})|^2 \right]=0,\label{equ:cal-en}
\end{align}
where
\begin{align}
\vartriangle\psi=\frac{\partial^2\psi}{\partial x^2}+\frac{\partial^2\psi}{\partial y^2},~~\vartriangle B_z=\frac{\partial^2 B_z}{\partial x^2}+\frac{\partial^2 B_z}{\partial y^2},
\end{align}
and $\rho$ and $T$ denote the density and the temperature; $v_x, v_y, v_z$ represent the $x-$component, $y-$component and $z-$component of the velocity, respectively; $\gamma$ is the polytropic index, which is selected to be $5/3$ in our simulation; $g$ is the normalized gravity; $\eta$ is the resistivity. Here $\beta_0=2\mu_0\rho_0RT_0L_0^2/\psi_0^2=0.1$ is the characteristic ratio of the gas pressure to the magnetic pressure, where $\rho_0=3.34\times10^{-13}\mathrm{~kg~m^{-3}}$, $T_0=10^6\mathrm{~K}$, $L_0=10^7\mathrm{~m}$, and $\psi_0=3.73\times10^3\mathrm{~Wb~m^{-1}}$ are the characteristic values of density, temperature, length and magnetic flux function, respectively, which are also the calculating units in the simulation. The characteristic values of other quantities are $v_0=128.57$ km s$^{-1}$, $t_0=77.8$ s, $B_0=3.37\times10^{-4}$ T, $g_0=1.65\times10^3$ m s$^{-2}$. The numerical domain is $0<x<200$ Mm, $0<y<300$ Mm, and discretized into 400$\times$600 uniform meshes with grid spacing $\vartriangle x=\vartriangle y=0.5$ Mm. Symmetric boundary condition is used for the left side ($x=0$). The radiation and the heat conduction in the energy equation are neglected. 
\par
In order to investigate the influence of flux feeding on flux rope systems, we must first construct a typical coronal flux rope system, and then realize the flux feeding process in simulations. Here we select a partially open bipolar field, with a negative and a positive surface magnetic charges located at the photosphere within $-b<x<-a$ and $a<x<b$, respectively, as the background field, which can be obtained by the complex variable method \citep[e.g.,][]{Hu1995,Zhang2017a}.
The background magnetic field can be cast in the complex variable form
\begin{align}
f(\omega)\equiv B_x-iB_y=\frac{(\omega+iy_N)^{1/2}(\omega-iy_N)^{1/2}}{F(a,b,y_N)}\mathrm{ln}\left( \frac{\omega^2-a^2}{\omega^2-b^2}\right),
\end{align}
where $\omega=x+iy$, and
\begin{align}
\nonumber &F(a,b,y_N)=\frac{1}{b-a}\int_a^b(x^2+y_N^2)^{1/2}dx=\frac{1}{2(b-a)}\times\\ &\left[b(b^2+y_N^2)^{1/2}-a(a^2+y_N^2)^{1/2}+y_N^2\mathrm{ln}\left(\frac{b+(b^2+y_N^2)^{1/2}}{a+(a^2+y_N^2)^{1/2}} \right)\right].
\end{align}
Here $a=30$ Mm, $b=40$ Mm, and ($y=y_N=60.6$ Mm, $x=0$) is the position of the neutral point of the partially open bipolar field. The neutral current sheet of the background field is located at $(x=0,~y\geq y_N)$. The width of the surface magnetic charges is $w=b-a=10$ Mm, and the distance between them is $d=2a=60$ Mm. The magnetic flux function could then be calculated by:
\begin{align}
\psi(x,y)=\mathrm{Im}\left\lbrace\int f(\omega)d\omega \right\rbrace,\label{equ:integral}
\end{align}
and the flux function at the lower base is
\begin{equation}
 \psi_i(x,0) = \left\{
              \begin{array}{ll}
              {\psi_c}, &{|x|<a}\\
              {\psi_c F(|x|,b,y_N)/F(a,b,y_N)}, &{a\leqslant|x|\leqslant b}\\
              {0}, &{|x|>b}
              \end{array}  
         \right.\label{equ:fluxb}
\end{equation}
where $\psi_c=\pi\psi_0$; the flux function at the neutral point $y=y_N$ is
\begin{align}
\psi_N=\frac{\pi(b^2-a^2)}{2F(a,b,y_N)}.\label{equ:fluxc}
\end{align}
The background field is potential everywhere except along the neutral current sheet and at the lower base. For potential magnetic fields, $f(\omega)=B_x-iB_y$ satisfies the Cauchy$-$Riemann condition, so that the integral in Equation \ref{equ:integral} is independent of the integration path blue in as far as the integration path does not touch the neutral current sheet and the lower base. The flux function along the neutral current sheet of the background partially open bipolar field is a constant, which is given by Equation \ref{equ:fluxc}; the flux function at the lower base is given by Equation \ref{equ:fluxb}. With the flux function calculated above, and let $B_z$ equals 0 in the background field, the configuration of the background field is obtained, as shown in \fig{fig:feeding}(a). The reconnection in the current sheet of the background field is prohibited by the method introduced in \cite{Hu2003a}. The initial corona is isothermal and static with
\begin{align}
T_c\equiv T(0,x,y)=1\times10^6 ~\mathrm{K},\ \  \rho_c\equiv\rho(0,x,y)=\rho_0\mathrm{e}^{-gy}.\label{equ:rhot}
\end{align}
As mentioned above, symmetric condition is used at the left boundary. Except during the emergence of the small rope, the lower boundary is fixed: the flux function $\psi$ is fixed at $\psi_i$ given by Equation \ref{equ:fluxb}; $B_z$ is fixed at 0; the velocity at the lower boundary is zero; the density and the temperature are fixed at their initial values, which are given by Equation \ref{equ:rhot}. The quantities at the right and top boundaries are evaluated by increment equivalent extrapolations \citep[e.g.][]{Hu2000a}:
\begin{align*}
U^{n+1}_{b}=U^{n+1}_{b-1}+U^{n}_{b}-U^{n}_{b-1}.
\end{align*}
Here $U$ represents the quantities (e.g. $\rho$, $\textbf{v}$, $\psi$); the superscript n and n+1 indicates the quantities at the current and the next time steps, respectively; $U_b$ represents the quantities at the boundary, and $U_{b-1}$ the quantities at the location next to the boundary. The boundary quantities at the next time step, $U^{n+1}_{b}$, are then prescribed.
\par
With the initial and background conditions, equations (\ref{equ:cal-st}) to (\ref{equ:cal-en}) are simulated by the multi-step implicit scheme \citep{Hu1989a}. Starting from the background field, first by letting a flux rope emerge from the lower base, we obtain a flux rope system with the rope sticking to the photosphere; then adjust the axial and poloidal fluxes of the rope to $\Phi_{z0}=9.31\times10^{19}~\mathrm{Mx}$ and $\Phi_{p0}=1.49\times10^{10}~\mathrm{Mx}~\mathrm{cm}^{-1}$, respectively, and let the rope system relax to a stable equilibrium state. The relaxation is achieved by letting the rope system evolve for a long enough time, during which the fluxes of the rope are fixed at $\Phi_{z0}$ and $\Phi_{p0}$; as a result of the numerical diffusion in the simulation, the rope system eventually reaches an equilibrium state. The final equilibrium state is just the initial state of our simulation (as shown in \fig{fig:feeding}(b)), and the rope with $\Phi_z=\Phi_{z0}$ and $\Phi_p=\Phi_{p0}$ in this state is the so-called major flux rope. This flux rope system is in a bald patch separatrix configuration. Note that the radius of the flux rope is finite here, i.e., there is no constraint on the ratio of the radius to, e.g., the characteristic photospheric length, so that the initial state could not be derived by analytical methods but could only be obtained by numerical procedures. 
\par

%\bibliographystyle{apj}
%\bibliography{feeding}

\begin{thebibliography}{}
\expandafter\ifx\csname natexlab\endcsname\relax\def\natexlab#1{#1}\fi

\bibitem[{{Antiochos} {et~al.}(1999){Antiochos}, {DeVore}, \&
  {Klimchuk}}]{Antiochos1999a}
{Antiochos}, S.~K., {DeVore}, C.~R., \& {Klimchuk}, J.~A. 1999, \apj, 510, 485

\bibitem[{{Archontis} \& {Hood}(2008)}]{Archontis2008b}
{Archontis}, V., \& {Hood}, A.~W. 2008, \apjl, 674, L113

\bibitem[{{Aulanier} {et~al.}(2010){Aulanier}, {T{\"o}r{\"o}k}, {D{\'e}moulin},
  \& {DeLuca}}]{Aulanier2010a}
{Aulanier}, G., {T{\"o}r{\"o}k}, T., {D{\'e}moulin}, P., \& {DeLuca}, E.~E.
  2010, \apj, 708, 314

\bibitem[{{Berger} {et~al.}(2010){Berger}, {Slater}, {Hurlburt}, {Shine},
  {Tarbell}, {Title}, {Lites}, {Okamoto}, {Ichimoto}, {Katsukawa}, {Magara},
  {Suematsu}, \& {Shimizu}}]{Berger2010}
{Berger}, T.~E., {Slater}, G., {Hurlburt}, N., {et~al.} 2010, \apj, 716, 1288

\bibitem[{{Bobra} {et~al.}(2008){Bobra}, {van Ballegooijen}, \&
  {DeLuca}}]{Bobra2008}
{Bobra}, M.~G., {van Ballegooijen}, A.~A., \& {DeLuca}, E.~E. 2008, \apj, 672,
  1209

\bibitem[{{Chen} \& {Shibata}(2000)}]{Chen2000a}
{Chen}, P.~F., \& {Shibata}, K. 2000, \apj, 545, 524

\bibitem[{{Chen} {et~al.}(2007){Chen}, {Hu}, \& {Sun}}]{Chen2007a}
{Chen}, Y., {Hu}, Y.~Q., \& {Sun}, S.~J. 2007, \apj, 665, 1421

\bibitem[{{Cheng} {et~al.}(2014){Cheng}, {Ding}, {Zhang}, {Sun}, {Guo}, {Wang},
  {Kliem}, \& {Deng}}]{Cheng2014}
{Cheng}, X., {Ding}, M.~D., {Zhang}, J., {et~al.} 2014, \apj, 789, 93

\bibitem[{{D{\'e}moulin} \& {Aulanier}(2010)}]{Demoulin2010a}
{D{\'e}moulin}, P., \& {Aulanier}, G. 2010, \apj, 718, 1388

\bibitem[{{Fan} \& {Gibson}(2007)}]{Fan2007a}
{Fan}, Y., \& {Gibson}, S.~E. 2007, \apj, 668, 1232

\bibitem[{{Forbes} \& {Isenberg}(1991)}]{Forbes1991a}
{Forbes}, T.~G., \& {Isenberg}, P.~A. 1991, \apj, 373, 294

\bibitem[{{Gopalswamy} {et~al.}(2017){Gopalswamy}, {Yashiro}, {Akiyama}, \&
  {Xie}}]{Gopalswamy2017}
{Gopalswamy}, N., {Yashiro}, S., {Akiyama}, S., \& {Xie}, H. 2017, \solphys,
  292, 65

\bibitem[{{Green} {et~al.}(2018){Green}, {T{\"o}r{\"o}k}, {Vr{\v s}nak},
  {Manchester}, \& {Veronig}}]{Green2018}
{Green}, L.~M., {T{\"o}r{\"o}k}, T., {Vr{\v s}nak}, B., {Manchester}, W., \&
  {Veronig}, A. 2018, \ssr, 214, 46

\bibitem[{{Guo} {et~al.}(2010){Guo}, {Ding}, {Schmieder}, {Li},
  {T{\"o}r{\"o}k}, \& {Wiegelmann}}]{Guo2010}
{Guo}, Y., {Ding}, M.~D., {Schmieder}, B., {et~al.} 2010, \apjl, 725, L38

\bibitem[{{Hu} {et~al.}(2015){Hu}, {Qiu}, \& {Krucker}}]{Hu2015}
{Hu}, Q., {Qiu}, J., \& {Krucker}, S. 2015, Journal of Geophysical Research
  (Space Physics), 120, 5266

\bibitem[{{Hu}(1989)}]{Hu1989a}
{Hu}, Y.~Q. 1989, Journal of Computational Physics, 84, 441

\bibitem[{{Hu} {et~al.}(2003){Hu}, {Li}, \& {Xing}}]{Hu2003a}
{Hu}, Y.~Q., {Li}, G.~Q., \& {Xing}, X.~Y. 2003, Journal of Geophysical
  Research (Space Physics), 108, 1072

\bibitem[{{Hu} {et~al.}(1995){Hu}, {Li}, \& {Ai}}]{Hu1995}
{Hu}, Y.~Q., {Li}, X., \& {Ai}, G.~X. 1995, \apj, 451, 843

\bibitem[{{Hu} \& {Liu}(2000)}]{Hu2000a}
{Hu}, Y.~Q., \& {Liu}, W. 2000, \apj, 540, 1119

\bibitem[{{Inoue} {et~al.}(2015){Inoue}, {Hayashi}, {Magara}, {Choe}, \&
  {Park}}]{Inoue2015}
{Inoue}, S., {Hayashi}, K., {Magara}, T., {Choe}, G.~S., \& {Park}, Y.~D. 2015,
  \apj, 803, 73

\bibitem[{{Isenberg} {et~al.}(1993){Isenberg}, {Forbes}, \&
  {Demoulin}}]{Isenberg1993a}
{Isenberg}, P.~A., {Forbes}, T.~G., \& {Demoulin}, P. 1993, \apj, 417, 368

\bibitem[{{Jiang} {et~al.}(2018){Jiang}, {Feng}, \& {Hu}}]{Jiang2018}
{Jiang}, C., {Feng}, X., \& {Hu}, Q. 2018, \apj, 866, 96

\bibitem[{{Kliem} {et~al.}(2014{\natexlab{a}}){Kliem}, {Lin}, {Forbes},
  {Priest}, \& {T{\"o}r{\"o}k}}]{Kliem2014}
{Kliem}, B., {Lin}, J., {Forbes}, T.~G., {Priest}, E.~R., \& {T{\"o}r{\"o}k},
  T. 2014{\natexlab{a}}, \apj, 789, 46

\bibitem[{{Kliem} \& {T{\"o}r{\"o}k}(2006)}]{Kliem2006a}
{Kliem}, B., \& {T{\"o}r{\"o}k}, T. 2006, Physical Review Letters, 96, 255002

\bibitem[{{Kliem} {et~al.}(2014{\natexlab{b}}){Kliem}, {T{\"o}r{\"o}k},
  {Titov}, {Lionello}, {Linker}, {Liu}, {Liu}, \& {Wang}}]{Kliem2014a}
{Kliem}, B., {T{\"o}r{\"o}k}, T., {Titov}, V.~S., {et~al.} 2014{\natexlab{b}},
  \apj, 792, 107

\bibitem[{{Lin} \& {Forbes}(2000)}]{Lin2000a}
{Lin}, J., \& {Forbes}, T.~G. 2000, \jgr, 105, 2375

\bibitem[{{Lin} {et~al.}(2001){Lin}, {Forbes}, \& {Isenberg}}]{Lin2001a}
{Lin}, J., {Forbes}, T.~G., \& {Isenberg}, P.~A. 2001, \jgr, 106, 25053

\bibitem[{{Liu} {et~al.}(2012){Liu}, {Kliem}, {T{\"o}r{\"o}k}, {Liu}, {Titov},
  {Lionello}, {Linker}, \& {Wang}}]{Liu2012b}
{Liu}, R., {Kliem}, B., {T{\"o}r{\"o}k}, T., {et~al.} 2012, \apj, 756, 59

\bibitem[{{Longcope} \& {Forbes}(2014)}]{Longcope2014a}
{Longcope}, D.~W., \& {Forbes}, T.~G. 2014, \solphys, 289, 2091

\bibitem[{{Moore} {et~al.}(2001){Moore}, {Sterling}, {Hudson}, \&
  {Lemen}}]{Moore2001a}
{Moore}, R.~L., {Sterling}, A.~C., {Hudson}, H.~S., \& {Lemen}, J.~R. 2001,
  \apj, 552, 833

\bibitem[{{Qiu} {et~al.}(2007){Qiu}, {Hu}, {Howard}, \& {Yurchyshyn}}]{Qiu2007}
{Qiu}, J., {Hu}, Q., {Howard}, T.~A., \& {Yurchyshyn}, V.~B. 2007, \apj, 659,
  758

\bibitem[{{Romano} {et~al.}(2003){Romano}, {Contarino}, \&
  {Zuccarello}}]{Romano2003}
{Romano}, P., {Contarino}, L., \& {Zuccarello}, F. 2003, \solphys, 214, 313

\bibitem[{{Savcheva} {et~al.}(2012){Savcheva}, {van Ballegooijen}, \&
  {DeLuca}}]{Savcheva2012b}
{Savcheva}, A.~S., {van Ballegooijen}, A.~A., \& {DeLuca}, E.~E. 2012, \apj,
  744, 78

\bibitem[{{Shen} {et~al.}(2014){Shen}, {Shen}, {Zhang}, {Hess}, {Wang}, {Feng},
  {Cheng}, \& {Yang}}]{Shen2014}
{Shen}, F., {Shen}, C., {Zhang}, J., {et~al.} 2014, Journal of Geophysical
  Research (Space Physics), 119, 7128

\bibitem[{{Sterling} \& {Moore}(2004)}]{Sterling2004}
{Sterling}, A.~C., \& {Moore}, R.~L. 2004, \apj, 602, 1024

\bibitem[{{Su} {et~al.}(2011){Su}, {Surges}, {van Ballegooijen}, {DeLuca}, \&
  {Golub}}]{Su2011a}
{Su}, Y., {Surges}, V., {van Ballegooijen}, A., {DeLuca}, E., \& {Golub}, L.
  2011, \apj, 734, 53

\bibitem[{{Su} {et~al.}(2009){Su}, {van Ballegooijen}, {Lites}, {Deluca},
  {Golub}, {Grigis}, {Huang}, \& {Ji}}]{Su2009}
{Su}, Y., {van Ballegooijen}, A., {Lites}, B.~W., {et~al.} 2009, \apj, 691, 105

\bibitem[{{Sun} {et~al.}(2012){Sun}, {Hoeksema}, {Liu}, {Wiegelmann},
  {Hayashi}, {Chen}, \& {Thalmann}}]{Sun2012a}
{Sun}, X., {Hoeksema}, J.~T., {Liu}, Y., {et~al.} 2012, \apj, 748, 77

\bibitem[{{T{\"o}r{\"o}k} \& {Kliem}(2003)}]{Torok2003a}
{T{\"o}r{\"o}k}, T., \& {Kliem}, B. 2003, \aap, 406, 1043

\bibitem[{van Driel-Gesztelyi \& Green(2015)}]{vanDriel2015a}
van Driel-Gesztelyi, L., \& Green, L.~M. 2015, Living Reviews in Solar Physics,
  12, 1

\bibitem[{{Wang} {et~al.}(2017{\natexlab{a}}){Wang}, {Liu}, {Ahn}, {Xu},
  {Jing}, {Deng}, {Huang}, {Liu}, {Kusano}, {Fleishman}, {Gary}, \&
  {Cao}}]{Wang2017a}
{Wang}, H., {Liu}, C., {Ahn}, K., {et~al.} 2017{\natexlab{a}}, Nature
  Astronomy, 1, 0085

\bibitem[{{Wang} {et~al.}(2017{\natexlab{b}}){Wang}, {Liu}, {Wang}, {Hu},
  {Shen}, {Jiang}, \& {Zhu}}]{Wang2017}
{Wang}, W., {Liu}, R., {Wang}, Y., {et~al.} 2017{\natexlab{b}}, Nature
  Communications, 8, 1330

\bibitem[{{Wang} {et~al.}(2015){Wang}, {Zhou}, {Shen}, {Liu}, \&
  {Wang}}]{Wang2015a}
{Wang}, Y., {Zhou}, Z., {Shen}, C., {Liu}, R., \& {Wang}, S. 2015, Journal of
  Geophysical Research (Space Physics), 120, 1543

\bibitem[{{Yan} {et~al.}(2018){Yan}, {Wang}, {Pan}, {Kong}, {Xue}, {Yang},
  {Li}, \& {Feng}}]{Yan2018}
{Yan}, X.~L., {Wang}, J.~C., {Pan}, G.~M., {et~al.} 2018, \apj, 856, 79

\bibitem[{{Zhang} {et~al.}(2001){Zhang}, {Dere}, {Howard}, {Kundu}, \&
  {White}}]{Zhang2001}
{Zhang}, J., {Dere}, K.~P., {Howard}, R.~A., {Kundu}, M.~R., \& {White}, S.~M.
  2001, \apj, 559, 452

\bibitem[{{Zhang} {et~al.}(2014){Zhang}, {Liu}, {Wang}, {Shen}, {Liu}, {Liu},
  \& {Wang}}]{Zhang2014}
{Zhang}, Q., {Liu}, R., {Wang}, Y., {et~al.} 2014, \apj, 789, 133

\bibitem[{{Zhang} {et~al.}(2016){Zhang}, {Wang}, {Hu}, \& {Liu}}]{Zhang2016a}
{Zhang}, Q., {Wang}, Y., {Hu}, Y., \& {Liu}, R. 2016, \apj, 825, 109

\bibitem[{{Zhang} {et~al.}(2017{\natexlab{a}}){Zhang}, {Wang}, {Hu}, {Liu}, \&
  {Liu}}]{Zhang2017a}
{Zhang}, Q., {Wang}, Y., {Hu}, Y., {Liu}, R., \& {Liu}, J. 2017{\natexlab{a}},
  \apj, 835, 211

\bibitem[{{Zhang} {et~al.}(2017{\natexlab{b}}){Zhang}, {Wang}, {Hu}, {Liu},
  {Liu}, \& {Liu}}]{Zhang2017}
{Zhang}, Q., {Wang}, Y., {Hu}, Y., {et~al.} 2017{\natexlab{b}}, \apj, 851, 96

\bibitem[{{Zhuang} {et~al.}(2018){Zhuang}, {Hu}, {Wang}, {Zhang}, {Liu}, {Gou},
  \& {Shen}}]{Zhuang2018}
{Zhuang}, B., {Hu}, Y., {Wang}, Y., {et~al.} 2018, Journal of Geophysical
  Research (Space Physics), 123, 2513

\end{thebibliography}

\end{document}